\documentclass[submission,copyright,creativecommons]{eptcs}
\providecommand{\event}{FMAS 2021} 

\usepackage{afterpage}
\usepackage{amsfonts}
\usepackage{amsmath}
\usepackage{amssymb}
\usepackage{bbold}
\usepackage{blindtext}
\usepackage{bm}
\usepackage{booktabs}
\usepackage{breqn}
\usepackage{caption}
\usepackage{cite}
\usepackage{courier}
\usepackage{enumitem}
\usepackage{float}
\usepackage[T1]{fontenc}
\usepackage{gensymb}
\usepackage{graphicx}
\usepackage{lipsum}  
\usepackage{listings}
\usepackage{lmodern}
\usepackage[utf8]{inputenc}
\usepackage{mathtools}
\usepackage[final]{pdfpages}
\usepackage{seqsplit}
\usepackage[usestackEOL]{stackengine}
\usepackage{tabularx}
\usepackage{textcomp}
\usepackage{tikz}
\usepackage{url}
\usepackage{color}
\usepackage[export]{adjustbox}
\usepackage{svg}
\usepackage{hyperref}
\usepackage{float}

{
  \begin{center}
  \par\vspace{\baselineskip}\noindent
}%
{
\end{center}
\par\addvspace{\baselineskip}
}

{
  \begin{center}
}%
{
\end{center}
\par\addvspace{\baselineskip}
}

\newtheorem{innercustomgeneric}{\customgenericname}
\providecommand{\customgenericname}{}

\makeatletter
\newcommand\aepath[1]{
  \bgroup
    \ttfamily
    \ae@path#1\relax\@nil
  \egroup}
\def\ae@path#1#2\@nil{
  \def\ae@continue{}
  \detokenize{#1}\unskip\penalty\z@  
  \ifx\relax#2
  \else 
    \def\ae@continue{\ae@path#2\@nil}
  \fi
  \ae@continue}
\makeatother
\let\mytexttt\aepath

\definecolor{codegreen}{rgb}{0,0.6,0}
\definecolor{cream}{rgb}{1.0,0.99,0.82}
\definecolor{codegray}{rgb}{0.5,0.5,0.5}
\definecolor{cornsilk}{rgb}{1.0, 0.97, 0.86}
\definecolor{codepurple}{rgb}{0.58,0,0.82}
\definecolor{backcolour}{rgb}{0.95,0.95,0.92}
\definecolor{athena}{rgb}{0.50, 0.20, 0.70}

\lstdefinestyle{athena-list}{
    commentstyle=\color{codegreen},
    numberstyle=\tiny\color{codegray},
    stringstyle=\color{codepurple},
    basicstyle=\scriptsize\ttfamily,
    breakatwhitespace=false,         
    breaklines=true,
    keywordstyle=\bfseries\color{athena},
    alsoletter=!-,  
    morekeywords={assert,load,datatype,domain,declare,define,conclude,pick-any,assume,let},                 
    captionpos=b,                    
    keepspaces=true,     
	mathescape=true,
    frame = single,
    frameround=tttt,
    rulecolor=\color{red}, 
    numbersep=1pt,                  
    showstringspaces=false,
    showtabs=false,                  
    tabsize=2,
    escapechar=\@
}

\lstdefinestyle{pseudo-list}{
    backgroundcolor=\color{cornsilk},   
    commentstyle=\color{codegreen},
    numberstyle=\tiny\color{codegray},
    stringstyle=\color{codepurple},
    mathescape =true, 
    basicstyle=\footnotesize\ttfamily,
    breakatwhitespace=false,         
    breaklines=true,                 
    captionpos=b,                    
    keepspaces=true,                 
    frame = single, 
    numbers=none,                    
    numbersep=5pt,                  
    showspaces=false,                
    showstringspaces=false,
    showtabs=false,                  
    tabsize=2,
    keywordstyle=\bfseries,
    morekeywords={End, If, While, then, Else, break, do}
}

\lstdefinestyle{equation-list}{
    basicstyle=\footnotesize\ttfamily,
    breakatwhitespace=false,         
    breaklines=true,
    alsoletter=!-,  
    captionpos=b,                    
    keepspaces=true,     
	mathescape=true,
    frameround=tttt,
    numbersep=1pt,                  
    showstringspaces=false,
    showtabs=false,                  
    tabsize=2,
    escapechar=\@ 
}

\makeatletter
\newsavebox{\@brx}
\newcommand{\llangle}[1][]{\savebox{\@brx}{\(\m@th{#1\langle}\)}%
  \mathopen{\copy\@brx\kern-0.5\wd\@brx\usebox{\@brx}}}
\newcommand{\rrangle}[1][]{\savebox{\@brx}{\(\m@th{#1\rangle}\)}%
  \mathclose{\copy\@brx\kern-0.5\wd\@brx\usebox{\@brx}}}
\makeatother

\DeclareMathAlphabet{\mathpzc}{OT1}{pzc}{m}{it}

\newcommand{\exval}[1]{\mathbf{E}\left[#1\right]}

\newcommand{\etal}[0]{$et~al.$}

\title{Formal Guarantees of Timely Progress for Distributed Knowledge Propagation}
\author{
Saswata Paul \qquad\qquad Stacy Patterson \qquad\qquad Carlos Varela
\institute{Rensselaer Polytechnic Institute, Troy, New York, 12180, USA}
\email{\quad pauls4@rpi.edu \quad\qquad \{sep,cvarela\}@cs.rpi.edu}}

\def\titlerunning{Guarantees of Timely Progress for Knowledge Propagation}
\def\authorrunning{S. Paul, S. Patterson, \& C. Varela}

\begin{document}
\maketitle

\begin{abstract}

Autonomous air traffic management (ATM) operations for urban air mobility (UAM) will necessitate the use of distributed protocols for decentralized coordination between aircraft.
As UAM operations are time-critical, it will be imperative to have formal guarantees of progress for the distributed protocols used in ATM.
Under asynchronous settings, message transmission and processing delays are unbounded, making it impossible to provide deterministic bounds on the time required to make progress.
We present an approach for formally guaranteeing timely progress in a Two-Phase Acknowledge distributed knowledge propagation protocol by probabilistically modeling the delays using theories of the Multicopy Two-Hop Relay protocol and the M/M/1 queue system.  
The guarantee states a probabilistic upper bound to the time for progress as a function of the probabilities of the total transmission and processing delays being less than two given values.
We also showcase the development of a library of formal theories, that is tailored towards reasoning about timely progress in distributed protocols deployed in airborne networks, in the Athena proof assistant.

\end{abstract}

\section{Introduction}\label{intro}
The integration of \textit{uncrewed aircraft systems} (UAS) in the \textit{National Airspace System} (NAS) for \textit{urban air mobility} (UAM) operations will pose significant challenges for urban air traffic management in the near future.
The UAS will need to operate in highly congested uncontrolled urban airspaces to perform tasks ranging from package delivery to passenger transportation.
Under such circumstances, it will be a complex task to ensure \textit{safe separation}~\cite{brittain2018autonomous} among the aircraft.
The loss of safe separation can have catastrophic consequences such as \textit{near mid-air collisions}~(NMAC)~\cite{lee2013investigating} and \textit{wake-vortex induced rolls}~\cite{luckner2004hazard}.
Centralized techniques for air traffic management, that require either human \textit{air traffic controllers} or dedicated \textit{ground stations} for coordinating the safe operation of aircraft, will be insufficient as they are prone to human errors~\cite{wing2011} and failures, and are economically infeasible~\cite{balachandran2020decentralized}.
Therefore, there is a need for developing decentralized \emph{UAS traffic management}~(UTM) protocols and standards for future UAM operations that will allow aircraft to autonomously coordinate and maintain safe separation. 
Such autonomous techniques will be amenable to \textit{formal verification}, making them suitable for safety-critical UAM applications.

\textit{Formal methods}~\cite{woodcock2009formal} can be used for the rigorous verification of critical properties of autonomous UTM applications by using techniques like \emph{model checking} and \emph{theorem-proving}.
However, the operational conditions presented by UAM are expected to be highly dynamic in nature, making it infeasible to exhaustively model them. 
Under such circumstances, it is possible to probabilistically reason about these conditions to provide valuable non-deterministic guarantees that can be used by the participating aircraft to take important real-time operational decisions~\cite{paul-dddas-2020}.
These guarantees can be supported by pre-developed formal proofs that have been mechanically verified for correctness under some reasonable assumptions about the conditions. 

\begin{figure}[]
  \centering
\resizebox{\columnwidth}{!}{

\tikzset{every picture/.style={line width=0.75pt}} 

\begin{tikzpicture}[x=0.75pt,y=0.75pt,yscale=-1,xscale=1]

\draw  [color={rgb, 255:red, 74; green, 144; blue, 226 }  ,draw opacity=1 ][dash pattern={on 1.69pt off 2.76pt}][line width=1.5]  (202.41,145.82) .. controls (202.41,71.91) and (262.32,12) .. (336.23,12) .. controls (410.14,12) and (470.06,71.91) .. (470.06,145.82) .. controls (470.06,219.73) and (410.14,279.65) .. (336.23,279.65) .. controls (262.32,279.65) and (202.41,219.73) .. (202.41,145.82) -- cycle ;
\draw (102.03,40.72) node [rotate=-122.45] {\includegraphics[width=27.04pt,height=31.92pt]{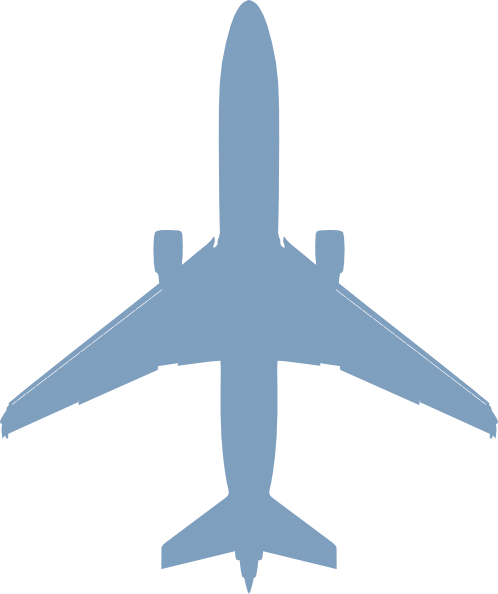}};
\draw (270.03,71.38) node [rotate=-137.65] {\includegraphics[width=27.04pt,height=31.92pt]{figures/aircraft_top.png}};
\draw (400.03,132.38) node [rotate=-81.04] {\includegraphics[width=27.04pt,height=31.92pt]{figures/aircraft_top.png}};
\draw (537.03,232.72) node [rotate=-283.77] {\includegraphics[width=27.04pt,height=31.92pt]{figures/aircraft_top.png}};
\draw  [dash pattern={on 4.5pt off 4.5pt}]  (112.8,231.6) -- (101.19,55.44) ;
\draw [shift={(101.06,53.44)}, rotate = 446.23] [color={rgb, 255:red, 0; green, 0; blue, 0 }  ][line width=0.75]    (10.93,-3.29) .. controls (6.95,-1.4) and (3.31,-0.3) .. (0,0) .. controls (3.31,0.3) and (6.95,1.4) .. (10.93,3.29)   ;
\draw  [dash pattern={on 4.5pt off 4.5pt}]  (145.8,239.6) -- (522.06,239.67) ;
\draw [shift={(524.06,239.67)}, rotate = 180.01] [color={rgb, 255:red, 0; green, 0; blue, 0 }  ][line width=0.75]    (10.93,-3.29) .. controls (6.95,-1.4) and (3.31,-0.3) .. (0,0) .. controls (3.31,0.3) and (6.95,1.4) .. (10.93,3.29)   ;
\draw (277.03,260.72) node [rotate=-59.33] {\includegraphics[width=27.04pt,height=31.92pt]{figures/aircraft_top.png}};
\draw  [dash pattern={on 4.5pt off 4.5pt}]  (499.8,31.6) -- (432.72,51.04) ;
\draw [shift={(430.8,51.6)}, rotate = 343.84000000000003] [color={rgb, 255:red, 0; green, 0; blue, 0 }  ][line width=0.75]    (10.93,-3.29) .. controls (6.95,-1.4) and (3.31,-0.3) .. (0,0) .. controls (3.31,0.3) and (6.95,1.4) .. (10.93,3.29)   ;
\draw   (11.06,6.44) -- (588.06,6.44) -- (588.06,297.64) -- (11.06,297.64) -- cycle ;
\draw  [dash pattern={on 4.5pt off 4.5pt}]  (505.8,105.6) -- (291.04,75.72) ;
\draw [shift={(289.06,75.44)}, rotate = 367.91999999999996] [color={rgb, 255:red, 0; green, 0; blue, 0 }  ][line width=0.75]    (10.93,-3.29) .. controls (6.95,-1.4) and (3.31,-0.3) .. (0,0) .. controls (3.31,0.3) and (6.95,1.4) .. (10.93,3.29)   ;
\draw  [dash pattern={on 4.5pt off 4.5pt}]  (505.8,105.6) -- (425.74,126.1) ;
\draw [shift={(423.8,126.6)}, rotate = 345.64] [color={rgb, 255:red, 0; green, 0; blue, 0 }  ][line width=0.75]    (10.93,-3.29) .. controls (6.95,-1.4) and (3.31,-0.3) .. (0,0) .. controls (3.31,0.3) and (6.95,1.4) .. (10.93,3.29)   ;
\draw  [dash pattern={on 4.5pt off 4.5pt}]  (505.8,105.6) -- (307.44,244.45) ;
\draw [shift={(305.8,245.6)}, rotate = 325.01] [color={rgb, 255:red, 0; green, 0; blue, 0 }  ][line width=0.75]    (10.93,-3.29) .. controls (6.95,-1.4) and (3.31,-0.3) .. (0,0) .. controls (3.31,0.3) and (6.95,1.4) .. (10.93,3.29)   ;

\draw (64,231) node [anchor=north west][inner sep=0.75pt]   [align=left] {Candidates};
\draw (502,18) node [anchor=north west][inner sep=0.75pt]   [align=left] {Airspace};
\draw (510,100) node [anchor=north west][inner sep=0.75pt]   [align=left] {Owners};

\end{tikzpicture}

}
  \caption{Aircraft trying to coordinate flight through a common airspace using DAC (top view).}
  \label{fig:airpace}
\end{figure}

In \cite{paul-nfm-2021} we have presented an autonomous protocol for UTM called \textit{Decentralized Admission Control}~(DAC) in which aircraft connected through an asynchronous \textit{vehicle-2-vehicle}~(V2V)~\cite{molisch2009survey} network called the \textit{Internet-of-Planes}~(IoP)~\cite{paul-dddas-2020} can coordinate their use of a shared four-dimensional airspace.
In DAC, an airspace has a set of \emph{owner} aircraft which already have the authorization to carry out a fixed set of \emph{flight plans} through the airspace (Fig.~\ref{fig:airpace}). 
The set of flight plans corresponding to the owners is \emph{safe}, \emph{i.e.}, there is no possibility of NMACs between any two flight plans in the set.
One or more \emph{candidate} aircraft can use the knowledge of this set to compute a \emph{conflict-aware flight plan}~\cite{paul-dasc-2019} that is compatible with all the owners.
A candidate can then propose its conflict-aware flight plan to the owners to request authorization to use the airspace.
Since the candidates do not consider each other in their conflict-free proposals, the owners can only authorize a single candidate at a time. 
When a candidate is authorized, the set of owners changes to include this candidate, and all aircraft \textit{relevant} to the airspace must be informed about the new set of owners so that they can learn the new value.
Aircraft relevant to an airspace comprises of the set of new owners and the set of candidates expected to try to enter the airspace in the future (this includes the candidates who may have been previously denied authorization).
Therefore, after a candidate is authorized, a \emph{knowledge propagation protocol} is needed to propagate the knowledge of the new set of owners to all relevant aircraft.

In \cite{paul-dasc-2020} we have presented a \textit{Two-phase Acknowledge} knowledge propagation protocol~(TAP) that can be used for \textit{sufficiently propagating} the knowledge of the new set of owners $\phi$ of an airspace to attain a \textit{safe state of knowledge}~\cite{fagin2004reasoning} for DAC.  
For TAP, we have formally proven \textit{consistency}, which implies that TAP will correctly propagate the knowledge, and \textit{eventual progress}, which implies that eventually, the safe state will be attained.
Although a guarantee of eventual progress \textit{alone} is insufficient for time-critical UAM applications, as it does not state any bound on the time for successful propagation, it is a necessary precondition for providing formal guarantees of \textit{timely progress}, which can provide some useful time bounds.
The total \textit{clock time} that is required for successful propagation is dependent on three factors---message transmission delays, message processing delays, and the number of messages involved.  
In asynchronous settings message delays are unbounded, making it impossible to provide deterministic bounds on the total time that may be required for successful propagation.

In the absence of deterministic time bounds, it is possible to provide probabilistic bounds for the timely progress of distributed protocols by modeling the factors affecting total time as stochastic processes.
In this paper, we formulate \textit{probabilistic timely progress guarantees} for TAP (assuming eventual progress) by using theories apposite to \emph{low-altitude platforms}~(LAP)~\cite{cao2018airborne} of airborne communication such as \emph{vehicular ad hoc networks}~(VANET)~\cite{hamdi2020review}.
Particularly, we formalize the theory of the \emph{Multicopy Two-Hop Relay}~(MTR) protocol~\cite{liu2013delivery} to model message transmission delays in VANETs and the \textit{M/M/1 queue system}~\cite{lipsky2009} to model message processing delays.
We then use the probabilistic bounds on these delays to provide sufficiently high-level guarantees of timely progress that are useful for UAM applications.
\textit{E.g.,} if a timely progress guarantee states that \textit{``propagation will take at most 5 seconds with a probability 98\%"}, then an aircraft can choose to only compute flight plans that start after 5 seconds to ensure that the plans don't become obsolete by the time that they are successfully propagated.

We have formalized our theory for timely progress in TAP using the \textit{Athena proof assistant}~\cite{athena1, arkoudas2017fundamental}.
Athena provides a specification and proof language along with an interactive proof development environment.
It is based on \emph{many-sorted first-order logic}~\cite{manzano1996extensions} and it uses a \emph{natural deduction}~\cite{arkoudas2005simplifying} style of proofs.
Athena is \emph{sound}, \textit{i.e.}, any method that successfully executes produces a theorem that is guaranteed to be a logical consequence of the \emph{assumption base}. 
Athena also provides some helpful mechanisms to modularize and organize theory in a \emph{reusable} manner. 
Using Athena, we have developed a formal theory library that is tailored towards reasoning about timely progress properties of distributed protocols deployed in airborne networks. 
This library can be used for reasoning about time-critical decentralized UAM applications. 

Another important motivation behind developing a tailored library was to lay the foundation for a \emph{parameterized proof library} that can be used in tandem with the \emph{dynamic data-driven applications systems}~(DDDAS)~\cite{darema2004dynamic} paradigm for the runtime verification of autonomous distributed applications~\cite{paul-dddas-2020}.
Autonomous UAM applications like DAC are expected to operate in unpredictable dynamic environments which cannot be completely modeled. 
Therefore, formal proofs that are developed in the pre-deployment stages may cease to be useful at runtime if the operational conditions assumed for the proofs do not correspond to the actual operational conditions. 
In the DDDAS-assisted approach, formal proofs can be parameterized by runtime-observable parameters about the operational conditions of distributed protocols and it is possible to augment the proofs in real-time using a \emph{formal DDDAS feedback loop}~\cite{paul-dddas-2020}.
A parameterized proof library will make it possible to provide formal guarantees that can be augmented with appropriate real-time parameters while retaining the formal rigor.
This will allow the development of highly adaptive autonomous distributed UAM applications that will be capable of adapting to the formal guarantees that are actually valid at runtime.

The main contributions of this paper are:
\begin{itemize}
    \item we formalize the theory of the Multicopy Two-Hop Relay (MTR) protocol to model message transmission delays between aircraft,
    
    \item we formalize the theory of the M/M/1 queue system to model message processing delays in each aircraft,
    
    \item we use the theory of MTR protocol and M/M/1 queue to provide machine-checkable probabilistic guarantees of timely progress for TAP, and
    
    \item we showcase the development of a formal proof library in Athena tailored towards timely progress guarantees of distributed protocols in VANETs.
\end{itemize}

The paper is structured as follows:
Section~\ref{propagation} describes TAP;
Section~\ref{guarantees} presents our timely progress guarantee for TAP;
Section~\ref{formalization} showcases the development of the proof library in Athena;
Section~\ref{discuss} presents a discussion on our approach towards developing the formal guarantees; 
Section~\ref{related} presents prior work related to progress analysis and formalization of mathematical theories;
and Section~\ref{conc} concludes the paper with a discussion about future directions of work.

\section{The Two-Phase Acknowledge Protocol}\label{propagation}

For decentralized UAM operations, aircraft will need to have \textit{heightened knowledge} about an airspace to operate with a \textit{sense of safety}~\cite{paul-dasc-2020}.
Fagin~\etal~\cite{fagin2004reasoning} succinctly explain this concern with the following example -- \emph{``..even if all drivers in a society know the rules for following traffic lights and follow them, that is not enough to make a driver ``feel safe''. This is because unless a driver knows that everyone else knows the rules and will follow them, then the driver may consider it possible that some other driver, who does not know the rules, may run a red light.''}. 
The knowledge propagation protocol (TAP) for DAC ensures a state of knowledge in which every aircraft can \textit{feel safe} to operate.
In \textit{knowledge logic}~\cite{fagin2004reasoning}, the expression $k_i \phi$ represents that \emph{an agent $i$ knows the fact $\phi$}, and $E \phi$ represents that \emph{all agents in the system know $\phi$}.
The expression $E E \phi$ (or $E^2 \phi$) represents a higher state of knowledge than $E \phi$ and implies that every agent knows two facts---$\phi$ and $E\phi$. 
In the context of DAC, $\phi$ represents the set of new owners after a candidate has been admitted and $E^2\phi$ represents the required safe state of knowledge which implies that all aircraft relevant to the airspace will know that all other aircraft know about the same new set of owners.
Aircraft relevant to an airspace comprises of the set of new owners and any known candidate expected to try to enter the airspace in the future.
This allows any future candidate to use the knowledge of $\phi$ to compute a correct conflict-aware plan.

\subsection{The Protocol}

TAP considers an asynchronous, non-Byzantine system model in which agents may fail temporarily and message delivery is reliable~\cite{gonczy2007modeling}.
There is a non-empty set of \textit{propagators}, and a logically separate non-empty set of \textit{replicas}.
All propagators know the set of replicas and the same value $\phi$.
$E^2 \phi$, in the context of our protocol, implies that \emph{all replicas know that all other replicas know $\phi$}.
The goal of every propagator is to propagate $\phi$ among all the replicas and eventually learn that $E^2 \phi$ has been achieved.
Propagators and replicas are logical abstractions and may be functionally implemented by the same physical node (\emph{e.g.,} an aircraft) simultaneously. 
For DAC, the set of aircraft which know about the new set of owners acts as the set of propagators and the set of all aircraft relevant to the airspace acts as the set of replicas. 

From the perspective of a propagator, TAP proceeds in two consecutive phases:
\begin{itemize}
\item \textbf{Phase 1}
\begin{enumerate}[label={(\alph*)}]
\item The propagator sends a \textit{learn} message with a value $\phi$ to each replica. 
 
\item A replica learns a value $\phi$ if and only if it receives a \textit{learn} message with the value $\phi$ from the propagator and it replies to the propagator with a \textit{learnt} message if and only if it has learned a value $\phi$.
\end{enumerate}

\item \textbf{Phase 2}
\begin{enumerate}[label={(\alph*)}]

\item The propagator sends an \textit{all-know} message to each replica if and only if it has received \textit{learnt} messages from all replicas.

\item A replica learns that all replicas know the value $\phi$ if and only if it receives an \textit{all-know} message from the propagator and it replies to the propagator with an \textit{acknowledgment} message if and only if it has learned that all replicas know the value $\phi$.

\item The propagator learns $E^2\phi$ has been achieved if and only if it receives \textit{acknowledgement} messages from all replicas. 
\end{enumerate}
\end{itemize}

\subsection{Progress in TAP}
We have formally proven in \cite{paul-dasc-2020} that under our system model, TAP will make eventual progress under some suitable conditions.
If eventual progress is guaranteed for TAP, then the total time taken for successful propagation is dependent on the following:
\begin{enumerate}
    \item message transmission delays,
    \item message processing delays, and
    \item the total number of messages involved.
\end{enumerate}
From the perspective of a propagator, successful propagation involves a deterministic number of messages.
For each replica, $4$ messages are required, so for $R$ replicas, the total number of messages required for successful propagation is $4 \times R$.
However, in asynchronous settings, transmission and processing delays are unbounded.
Therefore, even with a deterministic number of messages, it is impossible to provide a deterministic bound on the time for progress.

\section{Probabilistic Guarantees of Timely Progress}\label{guarantees}

In the absence of deterministic solutions, it is possible to employ a probabilistic approach for providing formal guarantees of timely progress.
This can be done by probabilistically reasoning about the message delays using either theoretical models or data-driven models. 
In contrast to theoretical models that are based on analytical results about system characteristics, data-driven results are usually based on real-time or historical statistical observations about the system.
This paper focuses on using theoretical models for reasoning about timely progress.

For UAM applications, it is desirable to choose theoretical models that are appropriate for airborne networks like the IoP.
Through the IoP, aircraft should be able to not only communicate their own information but also relay received information from other aircraft using a \textit{multi-hop ad~hoc} approach~\cite{wang2006fundamental}.
Moreover, an aircraft can be considered to be a single \textit{server} where messages arrive, wait for some time until they are processed, and take some time to be processed.
This makes it appropriate to use \textit{queueing theory}~\cite{lipsky2009} to reason about message processing delays. 
The M/M/1 system is an elementary queue system that consists of a single server that processes all messages which are received, making it a reasonable choice for modeling message processing with respect to a single aircraft.
M/M/1 system also provides some useful analytical properties that make it possible to formally reason about the message processing delays.  
Therefore, to formally derive timely progress guarantees for TAP, we assume the following:
\begin{itemize}
    \item all aircraft use the MTR protocol for message transmission between each other,
    \item each aircraft independently implements an M/M/1 queue to process the messages that it receives,
    \item the transmission delays of the messages are \emph{independent and identically distributed} (i.i.d.),
    \item the processing delays of the messages are i.i.d., and 
    \item the transmission delays are independent of the processing delays.
\end{itemize}

From the perspective of a propagator, if eventual progress is guaranteed, then in the worst-case, when there is no concurrency in message transmission and processing, the total time ($T_S$) taken for successful propagation can be calculated by using the total number of messages that are involved ($N_M$), and the transmission delay ($T_{D_m}$) and the processing delay ($T_{P_m}$) of each message $m$.
Since we know that $N_M$ has a deterministic value ($4 \times R$) for TAP, the total time for the worst-case scenario can, therefore, be obtained using Eq.~\ref{e0}.
\begin{equation}\label{e0}
    T_S = \sum_{m=1}^{N_M} T_{D_m} + \sum_{m=1}^{N_M} T_{P_m}
\end{equation}

Now, to derive a probabilistic bound on $T_S$, we first need to probabilistically model $T_{D_m}$ and $T_{P_m}$ for every message $m$ involved in progress.

\subsection{Modeling the Message Transmission Delays}
\emph{Two-hop relaying}, for data transmission between a \textit{source} and a \textit{destination} when the two nodes are not within \textit{transmission range}, has been proposed to be an efficient mode of communication in \textit{mobile ad hoc networks}~(MANET)~\cite{grossglauser2002mobility}.
Relaying has also been considered to be appropriate for airborne networks like the IoP where two communicating aircraft may not always be within direct \emph{transmission range}~\cite{wang2006fundamental}.
Therefore, to ensure that our assumptions about message transmission delays in airborne networks are consistent with prior work in the literature, we use the theory of the multicopy two-hop relay~(MTR) protocol.
For this, we base our assumptions on the description of the MTR protocol presented by Al Hanbali \textit{et al.}~\cite{al2007simple}.

In the basic two-hop relay protocol, there are a set of $M+1$ mobile nodes whose trajectories are i.i.d.~\cite{grossglauser2002mobility}.
If a \textit{source} node wants to transfer a message to a \textit{destination} node, it can either transmit it directly to the destination, if the destination is within its transmission range, or, it can transmit copies of the message to one or more \textit{relay} nodes.
A relay node can transmit a copy of a message directly to the destination node if the two are within the transmission range of the relay node.
A relay node, however, cannot transmit a copy of a message to another relay node.
Each message, therefore, makes a maximum of two hops---it is either transmitted directly from the source to the destination, or it is transmitted through one intermediate relay node.

In the MTR protocol~\cite{al2007simple}, transmission delay $T_{D_m}$ is the time taken for the destination to receive a message $m$ or a copy of $m$ for the first time.
Two nodes \emph{meet} when they come within the transmission range of one another and \textit{inter-meeting time} is the time interval between two consecutive meetings of a given pair of nodes.
The inter-meeting times of all pairs of nodes are i.i.d. with the common \emph{cumulative distribution function} (CDF) $G(t)$.
The source can only transmit a message to a relay that does not already hold a copy.
Message transmission between two nodes within range is instantaneous.
Each copy of a message has a \textit{time-to-live}~(TTL), which is the time after which a relay has to drop an untransmitted copy.

To derive a probabilistic bound on the time taken to transmit a message using MTR, we make the following assumptions:
\begin{itemize}
    \item the TTLs for all messages are unrestricted, \emph{i.e.}, a message can be held by a relay node until it can transmit it,
        
    \item since delivery is instantaneous when the nodes meet and TTLs are unrestricted, the inter-meeting time between the source and the destination ($T_{sd}$) or between a relay $i$ and the destination ($T_{r_{i}d}$) also represents the time taken by the source or the relay to directly deliver a message to the destination,

    \begin{equation}\label{e9}
        T_{D_m} = \mbox{min}\big(\{T_{sd}, T_{r_{1}d}, ..., T_{r_{M-1}d}\}\big)
    \end{equation}

    \item the source node transmits the copy of a message to all $M-1$ relay nodes. 
    A message (or a copy) can, therefore, be delivered to the destination by the source or any of the relay nodes.
    Hence, the actual time taken to deliver the message will be the minimum of $\{T_{sd}, T_{r_{1}d}, ..., T_{r_{M-1}d}\}$ (Eq.~\ref{e9}), and

    \item the inter-meeting times of the mobile nodes are \emph{exponentially distributed} with the \emph{rate parameter} $\lambda_{MTR}$. 
    
\end{itemize}
By using the above assumptions and necessary theories from probability, random variables, and exponential distributions, we have formally derived the following probabilistic bound on the time taken to deliver a message $m$ using MTR:
\begin{equation}\label{e10}
    P\left(T_{D_m} \leq t\right)= 1-(1- (1-e^{-\lambda_{MTR} t}))^M
\end{equation}

Now, the derivative of the above expression conforms to the \emph{probability density function}~(PDF) of an exponential distribution with rate parameter $\lambda_{MTR} M$. 
Therefore, we can conclude that $T_{D_m}$ is exponentially distributed with a rate parameter $\lambda_{D_m} = \lambda_{MTR} M$.

\subsection{Modeling the Message Processing Delays}
\textit{Queueing theory}~\cite{bertsekas1992data} has been extensively used in the literature for modeling throughput in MANETs~\cite{kushwah2020multipath, wang2012cooperation, yin2004malb, wen2011analysis}.
An aircraft communicating using the IoP can be considered to be a single \textit{server} where messages arrive, wait some time in a queue until they are processed, and take some time to be processed.
Therefore, for modeling the message processing delays in the aircraft, we use theory of queues, particularly the \emph{M/M/1} queue system~\cite{lipsky2009}.
We choose the M/M/1 queue system because it provides elegant analytical properties for computing the total time $T_{P_m}$ required for a message $m$ to be processed.
This processing time includes the time spent in the queue and the time spent in actually processing the message.
An important property of $T_{P_m}$ in the M/M/1 queue system is that $T_{P_m}$ is exponentially distributed~\cite{bertsekas1992data}.

The M/M/1 queue system consists of a single server where customers arrive according to a \textit{Poisson process}~\cite{last2017lectures}. 
It has the following characteristics~\cite{bertsekas1992data}:
\begin{itemize}
        \item the \textit{interarrival times} of messages in the queue are exponentially distributed with a rate parameter $\lambda_a$,
        \item the \textit{service time} of messages, which is the time spent in actually processing a message, is exponentially distributed with a mean $1 / \mu_s$, 
        \item the queue is managed by a single server, and
        \item the number of message arrivals in an interval of length $\tau$ follows a \textit{Poisson distribution}~\cite{leon-garcia_probability_1994} with a parameter $\lambda_a\tau$.
\end{itemize}

To probabilistically model $T_{P_m}$, we use \textit{Little's theorem}~\cite{little1961proof}, which is an important fundamental result in queueing theory. 
If $N_q$ and $T_p$ represent the \textit{average number of messages} and the \textit{mean processing delay} (queueing delay + service delay) respectively\footnote{Note that $T_p = \exval{T_{P_m}}$ is the \emph{mean} or the \emph{expected value} of $T_{P_m}$.}, then Little's Theorem states the useful relationship conveyed in Eq.~\ref{e13}. 
\begin{equation}\label{e13}
N_q = \lambda_a T_p
\end{equation}
A proof of Little's Theorem has been presented in \cite{bertsekas1992data}, which we have formalized\footnote{We present the detailed proof described in \cite{bertsekas1992data} to clearly highlight the current deficiencies in our formalization.}.
The proof uses $N(\tau)$, $\alpha(\tau)$, and $T(i)$ to represent the number of messages in the system at time $\tau$, the number of messages that arrived in the interval $[0,\tau]$, and the average time spent in the system by message $i$ respectively.
The average arrival rate over the time period $[0,t]$ is then given by:
\begin{equation}\label{e14}
\lambda_t = \frac{\alpha(t)}{t}
\end{equation}
Similarly, the average processing delay of messages that arrived in the interval $[0,t]$ is given by:
\begin{equation}\label{e15}
T_{t}=\sum_{i=1}^{\alpha(t)} \frac{T(i)}{\alpha(t)}
\end{equation}
The average number of messages in the system in the interval $[0,t]$ is given by:
\begin{equation}\label{e16}
N_{t}=\frac{1}{t} \int_{0}^{t} N(\tau) d \tau
\end{equation}
Now, the graphical analysis of a FIFO system reveals the following~\cite{bertsekas1992data}:
\begin{equation}\label{e17}
\frac{1}{t} \int_{0}^{t} N(\tau) d \tau=\frac{1}{t} \sum_{i=1}^{\alpha(t)} T(i)
\end{equation}
Using Eq.~\ref{e14} to Eq.~\ref{e17}, it can be shown that:
\begin{equation}\label{e18}
N_t = \lambda_t T_t
\end{equation}
Assuming that the limits $N_q=\lim _{t \rightarrow \infty} N_{t}$, $\lambda_a=\lim _{t \rightarrow \infty} \lambda_{t}$, and $T_p=\lim _{t \rightarrow \infty} T_{t}$ exist, we can now obtain Eq.~\ref{e13} to prove Little's Theorem. 

Now, using the properties of the Poisson distribution, it has been shown in~\cite{bertsekas1992data} that:
\begin{equation}\label{e20}
    N_q = \frac{\lambda_a}{\mu_s - \lambda_a}
\end{equation}
Again, from Eq.~\ref{e20} and Eq.~\ref{e13}, we can obtain the following relationship:
\begin{equation}\label{e21}
    T_p=\frac{1}{\mu_s-\lambda_a}
\end{equation}
Finally, since $T_{P_m}$ is exponentially distributed, its rate parameter $\lambda_{P_m}$ can be obtained as\footnote{For an exponentially distributed random variable, the rate parameter is the reciprocal of the mean.}:
\begin{equation}\label{e22}
    \lambda_{P_m} = \frac{1}{\exval{T_{P_m}}} = \frac{1}{T_p} = \mu_s-\lambda_a 
\end{equation}


\subsection{Probabilistic Bound on the Worst-Case Time}

Using the rate parameters $\lambda_{D_m}$ and $\lambda_{P_m}$ for the exponentially distributed message transmission and processing delays, we can now provide a probabilistic bound on the worst-case time for progress $T_S$. For that, let
\begin{equation}\label{en1}
    T_D = \sum^{N_M}_{m=1} T_{D_m} \ \ \ \ \ \& \ \ \ \ \ T_P = \sum^{N_M}_{m=1} T_{P_m}
\end{equation}
From Eq.~\ref{e0} and Eq.~\ref{en1}, for any two real numbers $x$ and $y$, we can state:
\begin{equation}\label{en2}
    ((T_D \leq x) \land (T_P \leq y)) \implies (T_S \leq (x + y))
\end{equation}
Now, for two events, $A$ and $B$, the following statement holds~\cite{meester_slooten_2021}:
\begin{equation}\label{en3}
    (A \implies B) \implies (P(B) \geq P(A))
\end{equation}
Therefore, from Eq.~\ref{en2} and Eq.~\ref{en3}, we get: 
\begin{equation}\label{en4}
     P(T_S \leq (x + y)) \geq P((T_D \leq x) \land (T_P \leq y))
\end{equation}
Since the processing delays are independent of the transmission delays, by the \textit{product rule} of independent events~\cite{gallager2013stochastic} we have\footnote{The product rule states that for independent events $A$ and $B$, $P(A \land B) = P(A) \times P(B)$.}: 
\begin{equation}\label{en5}
     P((T_D \leq x) \land (T_P \leq y)) = P(T_D \leq x) \times P(T_P \leq y)
\end{equation}
From Eq.~\ref{en4} and Eq.~\ref{en5} we have: 
\begin{equation}\label{en6}
      P(T_S \leq (x + y)) \geq P(T_D \leq x) \times P(T_P \leq y)
\end{equation}

Now, both $T_D$ and $T_P$ are sums of i.i.d. exponential random variables. 
Therefore, $T_D$ and $T_P$ follow two different \emph{Erlang distributions}~\cite{leon-garcia_probability_1994}, each of which are parameterized by a \emph{shape parameter} (which is $N_M$ in both cases) and a rate parameter ($\lambda_{D_m}$ for $T_D$ and $\lambda_{P_m}$ for $T_P$). 
The CDF of an Erlang distribution with shape parameter $k$ and rate parameter $\lambda$ is given by:
\begin{equation}\label{en7}
      F_{\mbox{\tiny Er}}(t,k,\lambda) = 1-\sum_{n=0}^{k-1} \frac{1}{n !} e^{-\lambda t}(\lambda t)^{n}
\end{equation}
This can be used to compute the delay probabilities as follows:
\begin{equation}\label{en8}
      P(T_D \leq x) = 1-\sum_{n=0}^{N_M-1} \frac{1}{n !} e^{-\lambda_{D_m} x}(\lambda_{D_m} x)^{n} = F_{\mbox{\tiny Er}}(x,N_M,\lambda_{D_m})
\end{equation}

\begin{equation}\label{en9}
      P(T_P \leq y) = 1-\sum_{n=0}^{N_M-1} \frac{1}{n !} e^{-\lambda_{P_m} y}(\lambda_{P_m} y)^{n} = F_{\mbox{\tiny Er}}(y,N_M,\lambda_{P_m})
\end{equation}

From Eq.~\ref{en6}, Eq.~\ref{en8}, and Eq.~\ref{en9}, we can state the required bound on $T_S$:
\begin{equation}\label{en10}
      P(T_S \leq (x + y)) \geq  F_{\mbox{\tiny Er}}(x,N_M,\lambda_{D_m}) \times  F_{\mbox{\tiny Er}}(y,N_M,\lambda_{P_m})
\end{equation}
where $N_M = 4 \times R$, $\lambda_{D_m} = \lambda_{MTR} M$  and $\lambda_{P_m} = \mu_s-\lambda_a $.

We have mechanically verified all the equations described in this section in Athena by using theory from limits, distributions, real numbers, and other domains.
However, currently, we have not formalized the theory required to reason about Poisson distributions and graphical analysis, and, therefore, have taken Eq.~\ref{e17} and Eq.~\ref{e20} as conjectures for our formal proofs.
Both these equations are established results from queuing theory that have been borrowed from Bertsekas~\etal~\cite{bertsekas1992data} and we plan to prove them in future work.

\section{A Library of Reusable Formal Theories in Athena}\label{formalization}

For verifying the correctness of distributed protocols that can be used in autonomous UTM applications, it is oftentimes necessary to prove complex high-level properties about the protocols.
Formally proving the correctness of a high-level property using an interactive proof assistant like Athena requires access to a formalization of all the fundamental theories on which the high-level property is based.
\emph{E.g.,} for proving the correctness of Eq.~\ref{en10}, it is necessary to formalize theories from a wide range of domains such as probability, MTR, and queues.
With the increasing complexity of the high-level properties, it becomes very arduous and time-consuming to formalize such fundamental theories from scratch, thereby making the task of verification intractable. 
For this reason, we are developing a formal library in Athena that is tailored for reasoning about high-level progress properties of distributed protocols in airborne networks\footnote{Complete Athena code available at http://wcl.cs.rpi.edu/pilots/fvcafp}. 

Athena provides a language for "proof programming" which relies on a fundamental technique called \emph{method call}.
Method calls in Athena represent logical \emph{inference rules} that can accept arguments of arbitrary types and are higher-order~\cite{musser-varela-agere-2013}.
Successful evaluation of a method call results in a \emph{theorem}, which is then added to the \emph{assumption base}.
The assumption base is a set of sentences that have either been proven to be correct by using inference rules or asserted to be correct.
Athena provides the \emph{soundness} guarantee that any theorem that is proven will be a direct consequence of sentences already in the assumption base.

Athena allows the use of \emph{modules} that can be used to organize, partition, and encapsulate theories into multiple logically separate \emph{namespaces}.
Modules are dynamic in the sense that they can be extended at any time and additional content can be added to them using the \texttt{extend-module} command (Fig.~\ref{fig:module}).
This feature of Athena allows theories, which are logically separate from one another, to be categorized and structured for brevity.
It also makes it easy to generalize theories for use in different contexts by importing them as required.

\begin{figure}[!ht]
\begin{lstlisting}[style=athena-list]
# Module Prob being extended by module PrbIndRv
load "Athena_LibDDDAS/math/Prob/Prob.ath"
extend-module Prob {
	module PrbIndRv {
    # contents of module PrbIndRv
    ...
    }
\end{lstlisting}
\caption{Extending modules in Athena.}
\label{fig:module}
\end{figure}

We have used our Athena library to formalize and verify all the high-level properties and theorems mentioned in Section~\ref{guarantees}.
For this purpose, we have formalized theories necessary to reason about distributed protocols in airborne networks by adding constructs that can be used to specify the properties of interest.
\emph{E.g.}, we have created a domain of distributed protocols \texttt{DistProt} that has properties which represent the characteristics of the operational environment such as \texttt{dpTotTim} (the total time for progress), \texttt{msGetTd} (the message transmission delays) and \texttt{msGetTp} (the message processing delays).
The Athena code for Eq.~\ref{en4}, which uses these constructs to define the probabilistic relationship between the total processing time, the message transmission delays, and the message processing delays, is given in Fig.~\ref{fig:eq16}.

\begin{figure}[!ht]
\begin{lstlisting}[style=athena-list]
# The relationship between total time for progress and message delays
define THEOREM-P-TS>=P-TD&TP :=
(forall DP r1 r2 .
		((Prob.probE	(Prob.consE Prob.<= (dpTotTim DP) (r1 + r2)))
			>=
			(Prob.probE	
				(Prob.cons2E 
					(Prob.consE Prob.<= (Random.SUM (msGetTd (dpGetMsgs DP))) r1) 
					(Prob.consE Prob.<= (Random.SUM (msGetTp (dpGetMsgs DP))) r2)))))	
\end{lstlisting}
\caption{The representation of Eq.~\ref{en4} in Athena.}
\label{fig:eq16}
\end{figure}

Similarly to \texttt{DistProt}, we have specified a domain of networks called \texttt{Network}, a domain of network protocols called \texttt{PrtType} that can represent protocols like MTR, a domain of queues called \texttt{Queue} to represent queuing models like M/M/1, and other specialized theories to reason about distributed protocols implemented over ad hoc airborne networks.

\begin{figure}[!ht]
\begin{lstlisting}[style=athena-list]
# Expected value of message processing delays in M/M/1 queue
conclude THEOREM-mean-T-value
...
let{
    ...
	N_d := 	
		( (Dist.ratePar (Random.pdf (srvcTm Q))) # mu_s	
		  -
		  (Dist.ratePar (Random.pdf (cstArRat Q))) # lambda_a
		);		
    ...
}		  
(!chain [ T
		= (N * (1.0 / L)) [T=N-x-inv-L]
		= ((L / N_d) * (1.0 / L)) [N=L-by-N_d]
		= (1.0 / N_d) [conn-2-a-by-d-x-b-by-a]
		])
\end{lstlisting}
\caption{A snippet from the proof of Eq.~\ref{e21} in Athena.}
\label{fig:eq11_athena}
\end{figure}
Athena supports the use of \emph{tactics} for interactively guiding the proof development of higher-level proof obligations from lower-level facts in the assumption base.
There are built-in methods in Athena that can be used to implement tactics for both \emph{forward} and \emph{backward} reasoning.
\emph{E.g.,}~Fig.~\ref{fig:eq11_athena} shows the use of the built-in \texttt{chain} method for deriving the proof of Eq.~\ref{e21} using forward reasoning, where \texttt{T} represents $T_p$ and \texttt{N_d} represents $\mu_s - \lambda_a$.   
In the future, we aim to define specialized methods that can be used as tactics for reasoning about distributed protocols.

\begin{figure}[!ht]
\begin{lstlisting}[style=athena-list]
# The relationship between cdf and probability
define cdf-prob-conjecture :=
(forall x R . ((Random.cdf x R) = (probE (consE <= x R))))
\end{lstlisting}
\caption{The relationship between CDF and probability defined as a conjecture in Athena.}
\label{fig:cdf}
\end{figure}
One of our goals while developing the library in Athena was reusability. 
As the proofs of different independent high-level properties are often dependent on common fundamental theories, it becomes beneficial to design the common theories in a manner that makes them reusable in different contexts.
\emph{E.g.,} the statement \texttt{cdf-prob-conjecture} (Fig.~\ref{fig:cdf}), which defines that the CDF $G_X(x)$ of a random variable $X$ is the probability that $X\leq x$ for all reals $x$, has been used in various contexts throughout our formalization of the results presented in Section~\ref{guarantees} (Fig.~\ref{fig:reus1}).
\begin{figure}[!ht]
\begin{lstlisting}[style=athena-list]
# Proof of P(min[X1,X2,...] <= y) = 1 - (1- G(X))^N
conclude THEOREM-probability-MIN-<=-IID-RVS-Gx 
...
	conn-to-cdf-prob := (!uspec (!uspec cdf-prob-conjecture (Random.rvSetIdElmnt rvSet)) R)
...


# Probabilistic bound on customer delay for M/M/1 queue 
conclude THEOREM-cstDly-prob
...
	conn-2-cdf-prob-conjecture := (!uspec (!uspec Prob.cdf-prob-conjecture (cstDly Q)) r)
...
\end{lstlisting}
\caption{Some instances where the \texttt{cdf-prob-conjecture} has been reused.}
\label{fig:reus1}
\end{figure}

\begin{figure}[!ht]
  \centering
  \includegraphics[width=\columnwidth]{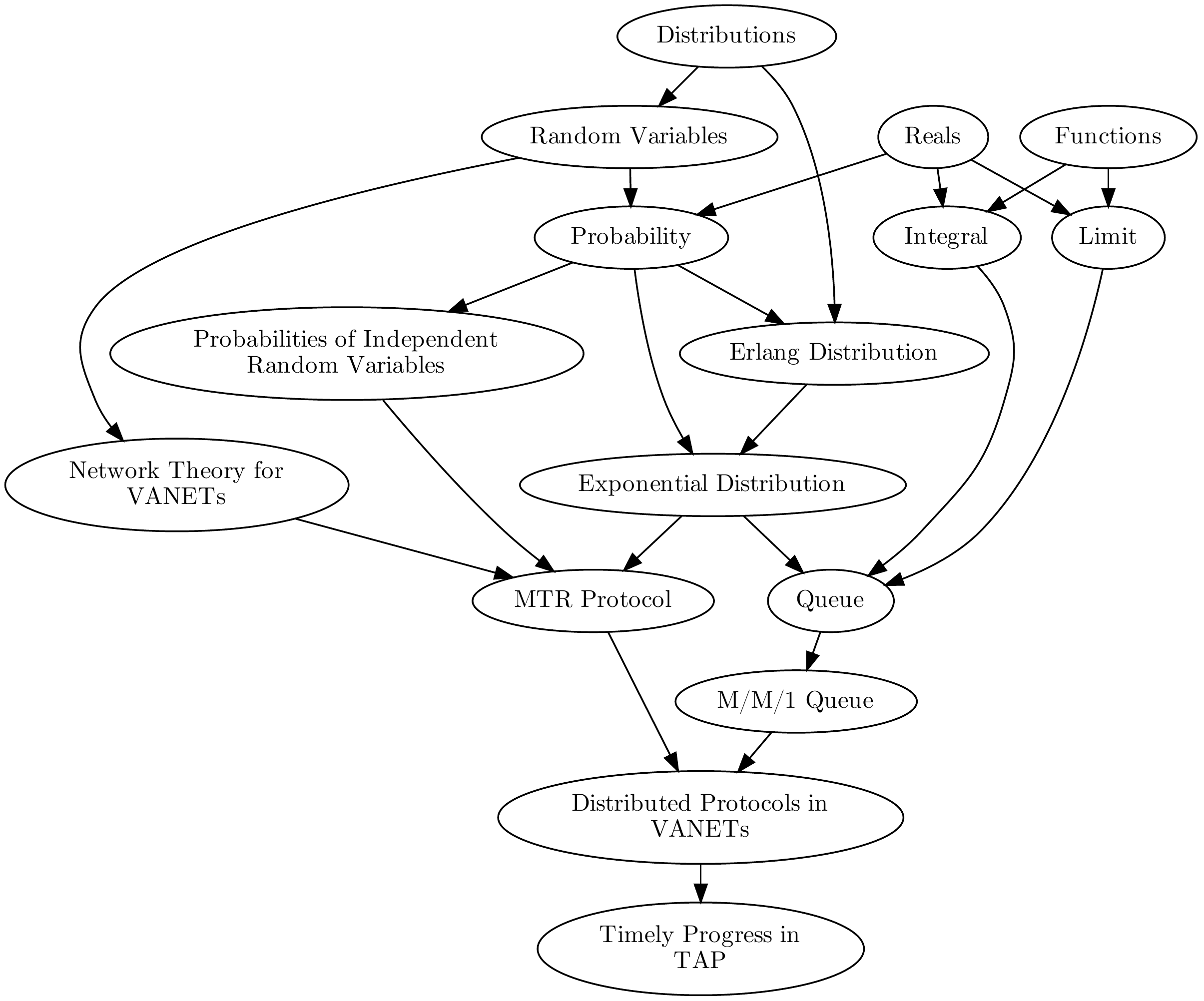}
  \caption{Current hierarchy of our extension to Athena's library for reasoning about distributed protocols in VANETs (does not depict Athena's built-in library).}
  \label{fig:root-hierarchy}
\end{figure}

At present, our extension to the Athena library consists of 71 axioms and 27 theorems, including the high-level properties of distributed protocols and other domains described in this paper.
It is composed of theories from three main domains---distributed systems, mathematics, and network theory.
Within each domain, there are theories from various related sub-domains. 
\emph{E.g.,} under mathematics, we have theories from probability, queues, random variables, functions, and distributions.
Fig.~\ref{fig:root-hierarchy} depicts the current logical hierarchy of our extension to the Athena library and the dependencies between the various modules representing the different logical domains and sub-domains.
As our primary goal was to develop high-level probabilistic progress properties of TAP that can be directly useful for UAM purposes, we adopted a \textit{top-down} approach of proof development where we only developed the lower-level theories that were required to support the higher-level properties of interest to us.
In most of the cases, we have tried to use only fundamental facts from the various domains as axioms, but there are instances in our formulation where we have currently taken some well-established results from the domains as conjectures that we aim to prove in future work~(\textit{e.g.,}~Eq.~\ref{e20}, \mytexttt{cdf-prob-conjecture}).

\section{Discussion}\label{discuss}

The availability of probabilistic guarantees of timely progress for autonomous distributed coordination protocols will allow the participating agents to make educated operational decisions.
\emph{E.g.}, if candidates for DAC know that it will take at most $x$ seconds to propagate the knowledge of their accepted plans with $99.99\%$ probability, then they can safely decide to compute plans that start after $x$ seconds, to ensure that the plans do not become stale.
Since the guarantees we have presented in Section~\ref{guarantees} are based on well-established theoretical models of VANETs, it makes them suitable for safety-critical airborne applications like DAC, where aircraft will need to coordinate over ad hoc networks in a decentralized manner.
Two-hop relaying has been proposed to be an efficient mode of communication in VANETs~\cite{grossglauser2002mobility} and since each aircraft will be independently responsible for processing all the messages sent to it, we believe that the proposed combination of the MTR protocol and the M/M/1 queue system is a reasonable approach for formally modeling communication between aircraft over the IoP.

Autonomous decentralized operations of mobile agents are complex in nature.
Such operations involve many interlinked aspects such as the mobility of the agents, the characteristics of the communication network, the internal computational capabilities of the agents, and the interaction between the agents.     
Therefore, to formally reason about such operations, it is necessary to holistically consider all these important aspects.
Achieving this in a machine-verifiable manner requires access to formal constructs, that are sufficiently expressive to correctly and completely specify such aspects, in some machine-readable formal language.
Developing such expressive formal constructs in a machine-readable language is a challenging task since it requires domain knowledge of all aspects of the system that need to be specified, strong knowledge of formal logic and reasoning techniques, experience and familiarity with the language in which the constructs are to be specified, and a significant amount of time and effort.  

In the absence of a pre-existing formal library that provides the necessary formal constructs to express some system, every time high-level properties of the system need to be verified, engineers have to develop the required formalizations from the ground up, often making the task of verification intractable. 
\emph{E.g.}, when we set out to specify timely progress properties for TAP using the theory of MTR protocol and M/M/1 queue system, we required theories in Athena to express constructs such as networks, mobility models, communication models, inter-agent interactions, queues, and distributed protocols, which were necessary to express the concepts presented in Section~\ref{guarantees}.     
For that reason, we developed most of the symbols, domains, and relationships necessary to comprehensibly express all the assumptions and specifications using the capabilities of Athena's many-sorted first-order logic.
We have tried to make the specifications reusable to different contexts so that further expansion of the library can be aided by using the existing specifications as building blocks rather than requiring to redevelop them for use in other contexts.
In our experience, efficiently designing reusable formal structures to express interconnected theories requires several iterations where the specifications need to be improved to be more general as new contexts for application are identified.
Another challenge is the meaningful modularization of the developed theories into appropriate categories to make it easy to import them independently in different contexts.
This is important because when theories are imported during proof development, they are automatically added to the assumption base.  
Well-defined modules allow theories to be imported only as needed without making the assumption base unnecessarily large.
This is an evolving project and we are certain that as more theory is added to the library, many of the formalizations in our current library will need to be redesigned in order to make them more general and better-organized.

During the development of our high-level formal guarantees of timely progress for TAP in Athena, we have relied on some theories which we have not formally verified in Athena, but have used them as conjectures for the proofs.
Nevertheless, we believe that this does not invalidate the results we have presented in the paper or diminish their importance in any way.
This is because all the conjectures we have used are well-established facts in the literature of the respective domains and have been informally proven beyond a reasonable doubt.
Since we were more interested in verifying the novel higher-level properties of timely progress for TAP that we have presented in Section~\ref{guarantees}, in the interest of time, we decided to take some of these well-established theories as foundations for developing our higher-level guarantees since our library currently lacks the necessary formal theory to mechanically verify them.
However, if time is not a concern, it should be possible to develop the proofs of these theories in Athena from just the fundamental axioms of the corresponding domains.

To the best of our knowledge, at the time of publication, our extension to the Athena library is the only library that provides a tailored formal foundation for reasoning about autonomous distributed coordination in airborne networks based on the theory of VANETs and queuing systems.
Therefore, we believe that our contribution will make the task of reasoning about properties of other distributed coordination protocols, which are implemented using MTR and the M/M/1 queue, in Athena easier in the future as a significant amount of reusable formal constructs to express such properties have already been developed.
Nonetheless, additional theories will be required to reason about distributed protocols involving a non-deterministic number of messages (\emph{e.g.}, the \emph{Synod consensus protocol}~\cite{paul-nfm-2021}).   
Our eventual goal is to expand the library by adding additional theories from distributions, queues, airborne networks, message sequences, and other necessary domains to support reasoning about a variety of distributed protocols that can be employed for autonomous decentralized UAM applications.

\section{Related Work}\label{related}
There is existing work in the literature on the informal analysis of timely progress of consensus.
Attiya~\etal~\cite{attiya1994bounds} provided lower-bounds on the time for progress in \textit{round-based}~\cite{delporte2007perfectly} consensus protocols.
Attiya~\etal~\cite{attiya2001time} analyzed the time complexity of solving distributed decision problems.   
Berman~\etal~\cite{berman1989towards} studied the number of message rounds involved in various consensus protocols.

Existing work on machine-verified guarantees has mainly analyzed eventual progress.
McMillan~\emph{et~al.}~\cite{mcmillan2018deductive} have proven eventual progress of \textit{Stoppable Paxos}~\cite{malkhi2008stoppable} in Ivy~\cite{padon2016ivy}.
Dragoi~\etal~\cite{druagoi2016psync} and Debrat~\etal~\cite{debrat2012verifying} have proven eventual progress in \emph{LastVoting}~\cite{charron2009heard}.
Hawblitzel~\emph{et~al.}~\cite{hawblitzel2015ironfleet,hawblitzel2017ironfleet} have proven eventual progress for a \textit{Multi-Paxos} implementation using Dafny~\cite{leino2010dafny}. 
In \cite{paul-nfm-2021}, we have verified eventual progress in the Synod consensus protocol using Athena.

Formalization of mathematical theories has also been presented in the literature. 
Hasan~\cite{hasan2008probabilistic} presented formalizations of statistical properties of discrete random variables in the \textit{HOL Theorem Prover}~\cite{gordon1989mechanizing}.
Hasan~\emph{et~al.}~have formalized properties of the standard uniform random variable~\cite{hasan2006formalization}, discrete and continuous random variables~\cite{hasan2008using, hasan2010formal}, tail distribution bounds~\cite{hasan2009formal}, and conditional probability~\cite{hasan2011reasoning}.
Mhamdi~\emph{et~al.}~\cite{mhamdi2010formalizationlebesque, mhamdi2013formalizationmeasure} have extended Hasan~\etal's work by formalizing \textit{measure theory} and the \textit{Lebesque integral} in HOL. 
HOL formalizations of the Poisson process, \textit{continuous chain Markov process}, and M/M/1 queue have been presented in \cite{chaouch2015formalization}.
Qasim~\cite{qasim2016formalization} has used Mhamdi~\etal's work to formalize the \textit{standard normal variable}.

In contrast to the existing work, we have developed formal theories tailored towards machine-verifiable timely progress guarantees for distributed protocols using models appropriate for airborne networks and have presented the timely progress guarantee of a knowledge propagation protocol that can be used for autonomous decentralized UAM applications.

\section{Conclusion and Future Work}\label{conc}
In this paper, we have presented a formal probabilistic guarantee of timely progress for the Two-phase Acknowledge knowledge propagation protocol by using theories from the Multi-copy Two-Hop Relay protocol and the M/M/1 queue system to reason about the non-deterministic message delays.
Since the progress guarantee provides useful probabilistic bounds on the time that may be required to make progress, it is more useful than a guarantee of eventual progress for time-critical UAM applications of the knowledge propagation protocol.
We have also showcased the development of a formal library in Athena for mechanically verifying the timely progress guarantee.
In the future, autonomous UAM applications will require participating aircraft to employ different types of distributed protocols for achieving different operational goals such as maintaining safe separation, autonomously coordinating operations through a shared airspace, etc. 
Our proof library has, therefore, been tailored to be reusable for reasoning about the timely progress of a variety of distributed protocols that can be used for such autonomous applications.

Currently, our formalizations have some high-level conjectures since our current library lacks the theories required to formally verify them.
In the future, we plan to formalize the necessary fundamental theories required to prove these conjectures as theorems.
The library also does not support the complete formalization of timely progress if the number of messages is non-deterministic.
Therefore, another potential direction of future work is to add additional theories to the library to allow reasoning about distributed protocols that involve a non-deterministic number of messages, such as the Synod consensus protocol.
In the future, we would also like to model other routing protocols suitable for VANETs in addition to MTR so that appropriate guarantees can be provided depending on actual implementations.

\hfill \break \noindent\textbf{Acknowledgment:} This research was partially supported by the National Science Foundation (NSF), Grant No. -- CNS-1816307 and the Air Force Office of Scientific Research (AFOSR), DDDAS Grant No. -- FA9550-19-1-0054.

\bibliographystyle{eptcs}
\bibliography{references}

\end{document}